\documentstyle[12pt,iopfts]{ioplppt}

\newcommand{\be}{\begin{equation}}
\newcommand{\ee}{\end  {equation}}
\newcommand{\beq}{\begin{eqnarray}}
\newcommand{\eeq}{\end{eqnarray}}
\newcommand{\pa}{\partial}
\newcommand{\half}{{1\over 2}}

\newtheorem{theorem}{Theorem}

\newcommand{\pat}{\partial_t}
\newcommand{\px}{\partial_x}
\newcommand{\crr}{\nonumber\\}

\begin{document}
\title{ Wilson loops from supergravity and string theory}

\author{ J. Sonnenschein}

\address{
 Raymond and Beverly Sackler Faculty of Exact Sciences\\
 School of Physics and Astronomy\\
 Tel Aviv University, Ramat Aviv, 69978, Israel}

\begin{abstract}
We present a theorem that determines the value of the Wilson loop
associated with a   Nambu-Goto action which generalizes the action of the 
$AdS_5\times S_5$ model. In particular we derive  sufficient conditions
for confining behavior.
We then apply this theorem to various string models. 
We go beyond the classical string picture by incorporating  quadratic quantum 
fluctuations. We show that
the bosonic determinant of $D_p$ branes with 16 supersymmetries yields
a Luscher term.
 We confirm that the free energy  associated with 
a BPS 
configuration of a single quark is free from divergences. 
We show that  unlike for a string in flat space time 
in the case of $AdS_5\times S_5$ the fermionic determinant does not
cancel the bosonic one. For a setup that corresponds to a confining gauge
theory the correction to the potential is attractive. 
We determine the form of the Wilson loop for actions that include non trivial 
$B_{\mu\nu}$ field. The issue of an exact determination of the value of the
stringy Wilson loop is discussed. 

Talk presented in string 99 Potsdam.
\end{abstract}

\section{ Classical Wilson loops - general results }
Wilson loops were derived from  the Nambu-Goto action
associated with the  $AdS_5\times S^5$ 
supergravity\cite{Mal2}\cite{Rey} and from several other related 
actions\cite{BISY1}-\cite{gppz}
We introduce here a space-time metric that unifies all these models and determine
the corresponding static potential\cite{KSS}. 

Consider a 10d space-time metric 
\be
 ds^2 = -G_{00}(s) dt^2 + G_{x_{||}x_{||}}(s) dx^2_{||} 
+ G_{ss}(s) ds^2
 + G_{x_{T}x_{T}}(s) dx^2_{T} 
\ee
where  
$x_{||}$ are $p$ space coordinates on a $D_p$ brane  and 
$s$ and  $x_T$ are the transverse coordinates. 
The corresponding Nambu-Goto action is 
$$
S_{NG} =\int d\sigma d\tau \sqrt{det[\pa_\alpha  x^\mu \pa_\beta x^\nu
G_{\mu\nu}]} \crr
$$
Upon using $\tau=t$ and  $\sigma=x$, where  $x$ is  one of the $x_{||}$
 coordinates,   the  action 
for a static configuration  reduces  to  
$$
S_{NG}= T \cdot \int dx\sqrt{ f^2(s(x)) + g^2(s(x)) (\pa_x s)^2} \crr
$$ 
where
\be
f^2(s(x))  \equiv  G_{00}(s(x))G_{x_{||}x_{||}}(s(x)) \qquad
g^2(s(x))  \equiv  G_{00}(s(x))G_{ss}(s(x)) 
\ee
and $T$ is the time interval. 
The equation of motion  (geodesic line) 
$$
\label{sofx}
\frac{d s}{d x} = \pm \frac{f(s)}{g(s)} \cdot
\frac{\sqrt{f^2(s)-f^2(s_0)}}{f(s_0)}
$$
For a static string configuration connecting ``Quarks" separated  
by a distance
$$
\label{lgeneral}
L = \int dx = 
2 \int_{s_0}^{s_1} \frac{g(s)}{f(s)} \frac{f(s_0)}{\sqrt{f^2(s)-f^2(s_0)}} ds 
$$
The NG action   
and   the corresponding  energy  $E={S_{NG}\over T}$ are 
divergent.  
The action  is renormalized  by subtracting  the quark masses\cite{Mal2}.
 For the $AdS_5\times S^5$ case it is equivalent to the procedure 
suggested by \cite{DGO}.
\be
m_q = \int_0^{s_1} g(s) ds
\ee
So that the renormalized quark anti-quark potential is   
\be
E  =  f(s_0) \cdot L  
 + 2 \int_{s_0}^{s_1} \frac{g(s)}{f(s)}  
( \sqrt{f^2(s)-f^2(s_0)} - f(s)  ) ds - 2 \int_0^{s_0} g(s) ds 
\ee 
The behavior of the potential is determined by  the following theorem\cite{KSS}. 
\begin{theorem}
\label{main}
Let $S_{NG}$ be  the NG action defined above, with functions $f(s),g(s)$ such
that:
\begin{enumerate}
\item \label{smoothf} $f(s)$ is analytic   for $0 < s < \infty$.
At $s = 0$, ( we take here that the minimum of $f$ is at $s=0$ ) its expansion is:
$$
\label{expansion}
f(s) = f(0) + a_k s^k + O(s^{k+1})
$$ 
with $k > 0 \;,\; a_k > 0$.
\item \label{smoothg} $g(s)$ is smooth for $0 < s < \infty$. At $s = 0$,
its expansion is:
$$
g(s) = b_j s^j + O(s^{j+1})
$$
with $j > -1 \;,\; b_j > 0$.
\item \label{positive} $f(s),g(s) \ge 0$ for $0 \le s < \infty$.
\item \label{increasing} $f'(s) > 0$ for $0 < s < \infty$.
\item \label{inftyint} $\int^\infty g(s)/f^2(s) ds < \infty$.

\end{enumerate}
Then for (large enough) $L$ there will be an even geodesic line asymptoting
from both sides to $s = \infty$, and $x = \pm L/2$.
The associated potential  is 
\begin{enumerate}
\item \label{conf}   if $f(0) > 0$, then
\begin{enumerate}
\item if $k = 2(j+1)$, 
\vskip 1cm
\begin{center}
\fbox{
$E = f(0) \cdot L -2 \kappa + 
                                    O((\log L)^\beta  e^{-\alpha L})$}
\end{center}
\item if $k > 2(j+1)$, 
\vskip 1cm
\begin{center}
\fbox{
$E = f(0) \cdot L -2 \kappa - d \cdot L^{-\frac{k+2(j+1)}{k-2(j+1)}} 
+ O(L^\gamma ) $.}
\end{center}
\end{enumerate} where $\gamma={-\frac{k+2(j+1)}{k-2(j+1)} - \frac{1}{k/2-j}}$
 and $\beta$ and $\kappa$, $\alpha$  $d$ and   $C_{n,m}$ are
positive constants determined by the string configuration.

In particular, there is  {\bf linear confinement}

\item \label{noconf} if $f(0) = 0$, then if $k > j+1$,
\begin{center}
\fbox{
$E = -d' \cdot L^{-\frac{j+1}{k-j-1}}
 + O(L^{\gamma'})$} 
\end{center}
where $\gamma'={-\frac{j+1}{k-j-1} 
- \frac{2k-j-1}{(2k-j)(k-j-1)}}$ and 
  $d'$ is a  coefficient determined by the classical configuration.

In particular,   there is  {\bf no  confinement}
\end{enumerate}
\end{theorem}

A detailed proof of this theorem is given  in \cite{KSS}.
As a consequence of this theorem it is straightforward that 
a sufficient 
 condition for confinement is 
if either of the two conditions is  obeyed:

(i) $f$ has a minimum at $s_{min}$
and $f(s_{min})\neq 0$ or

(ii) 
 $g$ diverges at $s_{div}$  and 
 $f(s_{div})\neq 0.$\\

\section{ Applications to various models}
\begin{tabular}{|l||c|c|}\hline

Model& Nambu-Goto Lagrangian & Energy \\ \hline\hline
 & &    \\
$AdS_5\times S^5\cite{Mal2}$ &$ \sqrt{U^4/R^4 + (U')^2}$ &
$- \frac{2\sqrt{2}{\pi}^{3/2}R^2}{\Gamma(\frac{1}{4})^4}  \cdot L^{-1} $
 \\
 & &    \\\hline
 & &    \\
non-conformal & &    \\
$D_p$ brane\cite{IMSY,BISY2}   & $ \sqrt{(U/R)^{7-p} + (U')^2}$ & 
$ -d' \cdot L^{-2/(5-p)} $   \\
(16 supersymmetries)  & &  $+O(l^{-2/(5-p) - 2(6-p)/(5-p)(7-p)})$  \\
 & &    \\\hline
 & &    \\
Pure YM  in 4d at&
 & $\sim L^{-1}(1-c(LT)^4)$ for $L<<L_c$ 
  \\
 finite temperature &$ \sqrt{(U/R)^4 (1-(U_T/U)^4) + (U')^2}$ &
  full screening   $ L>L_c$\\
\cite{BISY2,RTY}& &   \\\hline
 & &    \\
Dual model of  & &   \\
Pure YM  in 3d &$  \sqrt{(U/R)^4 + (U')^2 (1-(U_T/U)^4)^{-1}}$ & 
$ \frac{U_T^2}{2\pi R^2} \cdot L -2\kappa + O(\log l \; e^{-\alpha L})$   \\
\cite{Witads2,BISY2,DorOtt,GriOle} & &    \\\hline
 & &   \\
Dual model of  & &    \\
Pure YM  in 4d &$ \sqrt{(U/R)^3 + (U')^2 (1-(U_T/U)^3)^{-1}}$ &  
$\frac{U_T^{3/2}}{2\pi R^{3/2}} \cdot L -2\kappa + O(\log L \; e^{-\alpha L})$
  \\
& &    \\\hline

 & &    \\
Rotating $D_3$ brane\cite{CORT} & $ \sqrt{C} \sqrt{ \frac{U^6}{U_0^4} \Delta + (U')^2 \frac{U^2 \Delta}
               {1-a^4/U^4 - U_0^6/U^6}}$ &$4/3{U_T^2\over R^2}C L+...$    \\
& &    \\\hline
 & &    \\
$D_3+D_{-1}$ system\cite{LuiTse} &$ \sqrt{(U^4/R^4+q) + (U')^2(1+qR^4/U^4)}
$ & qL+...   \\              
 & &    \\\hline
 & &    \\
$MQCD$ system\cite{KSS1} &$  2 \sqrt{2\zeta} \sqrt{\cosh(s/R_{11})} \sqrt{1+s'^2}$ &
$E = 2 \sqrt{2\zeta} \cdot L -2\kappa $ 
    \\  
& &  $+ O(\log L \; e^{-1/\sqrt{2}R_{11} L})$  \\
 & &    \\\hline
 & &    \\
't Hooft loop\cite{BISY1,GroOog} 
&$\frac{1}{g^2_{YM}} \sqrt{(U/R)^3 (1-(U_T/U)^3)  + (U')^2 }$
&  full screening   \\
 & & of monopole pair    \\\hline
\end{tabular}

\section{Quantum fluctuations}
So far we have discussed Wilson loops from their correspondence to
certain classical string configuration. Now we write down  the machinery
to incorporate quantum fluctuations and present some preliminary results
about the QM determinant of some of the classical setups discussed above
\cite{KSSW}.

We start with introducing   quantum fluctuations around the classical 
bosonic configuration
$$
x^\mu(\sigma,\tau) = x^\mu_{cl}(\sigma,\tau) + \xi^\mu(\sigma,\tau)
$$
The quantum corrections  of the Wilson line is 
$$
	\langle W\rangle = {\rm e}^{-E_{cl}(L)T}~
	\int \prod_a d\xi_a \exp\left(-\int d^2 \sigma \sum_a \xi^a {\cal O}^a\xi^a \right)
$$
where $\xi^a$ are the fluctuations left after  gauge fixing. 
The corresponding  free energy is 
$$
F_B = -\log {\cal{Z}}_{(2)}= -\sum_a\half \log \det {\cal O}_a
$$

\subsection{ Gauge fixing}
In the classical treatment it is convenient to choose for the worldsheet
 coordinates $\tau=x^0$ and $\sigma=x$. 
In computing the quantum corrections it seems that there are several equivalent 
gauge fixings. One can still use the gauge of above, namely set $\xi_x=0$, or
fix $\sigma=u,\  \xi_u=0$ (we denote here $s$ by $u$) so that there are no
fluctuations in the  space-time metric. However, it turns out that those gauges
suffer from singularities at  the  minimum of the configuration $u_0$. 
A gauge that is free from those singularities is the ``normal coordinate
gauge" $\sigma=u_{cl}$ and the fluctuation 
in $x,u$ plane
is in the coordinate normal to  $u_{cl}$. 

\subsection{ General form of the bosonic determinant
}

In the $\sigma=u$ gauge and  after a change of variables the free energy 
is given by 
\be 
\label{FrEn}
       F_B = - \half \log {\det {\cal O}_{x}}
 - {(p-1)\over 2} \log {\det {\cal O}_{x_{II}}} 
- {(8-p)\over 2} \log {\det {\cal O}_{x_T}}
\ee
where 
\beq
\label {QuadOp} 
\hat{\cal O}_x &=& \left[\px\left((1-{f^2(u_0) \over f^2(u_{cl})})
\px\right) + {G_{xx}(u_{cl}) \over G_{tt}(u_{cl})} ({f^2(u_{cl}) \over f^2(u_0)}-1)
\pat^2 \right] \nonumber \\
\hat{\cal O}_{x_{II}} &=&  \left[\px\left({G_{y_i
y_i}(u_{cl}) \over G_{xx}(u_{cl})} \px\right) + {G_{y_i y_i}(u_{cl}) \over G_{tt}(u_{cl})} 
{f^2(u_{cl}) \over f^2(u_0)} \pat^2 \right]   \nonumber \\
\eeq
with $\hat{\cal O}= {2\over f(u_0)}{\cal O}$ and a similar expression
for $\hat{\cal O}_{x_T}$.
The boundary conditions are $\hat\xi(-L/2,t) = \hat\xi(L/2,t)=0$

\subsection{ Bosonic fluctuations in flat space-time }
Let us recall first the determinant in flat space-time.
The fluctuations  in this case are determined by the following action 
$$
	S_{(2)}=\half\int d\sigma d\tau
\sum_{i=1}^{D-2}\left[ 
	\left(\partial_\sigma\xi_i\right)^2 +
	\left(\partial_\tau\xi_i\right)^2
	\right]
$$
The  corresponding eigenvalues are 
$$
       E_{n,m} = ({{n \pi \over L}})^2 + ({{m \pi} \over T})^2
$$
and the free energy is given by
$$
       -\frac{2}{D-2} F_B =   \log \prod_{n m} E_{n,m} = 
 T {\pi \over 2L} \sum_n n + O(L)
$$

Regulating this result  using Riemann $\zeta$ 
function we find
that the  quantum correction to the linear quark anti-quark potential 
is

\begin{center}
\fbox{
$
      \Delta V(L) = - {1 \over T}  F_B 
= -(D-2) {\pi \over 24} \cdot {1 \over L}$
}
\end{center}

which is the so-called {\bf  L\"{u}scher term}\cite{Luscher}.

\subsection{General scalling  relation, and
 the $L$ dependence of $\Delta V$ ? 
}


Consider an operator of the form 
$$
{\cal O}[A,B] = A^2 F_t(v) \partial_t^2 + B^2 \partial_v (F_v(v) \partial_v)
$$
The correction  to the potential $V[A,B]$
 due to fluctuations determined by such an
operator  is
 $$V[A,B] = (B/A) \cdot V[1,1]$$
For  the operators that describe the  fluctuations 
associated with  metrics such that
\be
f(u)  =  a u^k \qquad
g(u)  =  b u^j
\ee
like for  instance  for the $D_p$ brane solution in the near horizon limit
 we find that 
$A^2 = b u_0^j \;,\; B^2 = \frac{a^2}{b} u_0^{2k-j-2}\rightarrow 
B/A = \frac{a}{b} u_0^{k-j-1}$
Therefore, the
potential is proportional to 
\begin{center}
\fbox{
 $B/A = \frac{a}{b} u_0^{k-j-1}
\rightarrow \Delta V\propto L^{-1}$
}
\end{center}

Thus, the quantum correction of the quark anti-quark potential  is of  L\"{u}scher 
type\cite{Luscher}for models of $D_p$ branes with 16 supersymmetries, in particular also the 
$AdS_5\times S^5$ model.  
\subsection { The fermionic fluctuations in flat space-time }

The NSR action of the type II superstring with a RR fields like 
on $AdS_5\times S^5$ is not known.
On the other hand
the  manifestly  space-time  supersymmetric Green Schwarz  
action was written down for the $AdS_5\times S^5$ case\cite{MatTse}
 To demonstrate the use of the GS action we start with the 
fermionic determinant in flat space-time

 The fermionic part of the $\kappa$ gauged fixed  GS-action is
$$
    S_F^{flat} = 2 i  \int d\sigma d \tau  \bar \psi  \Gamma^i \pa_i \psi 
$$
where $\psi$ is a Weyl-Majorana spinor, $\Gamma^i$ are 
the SO(1,9) gamma matrices, $i,j = 1,2$ and we explicitly considered a flat classical string.
Thus the fermionic operator is
$$ 
    \hat{\cal O}_F=D_F = \Gamma^i \pa_i \qquad
(\hat{\cal O}_F)^2=\Delta = \px^2 - \pat^2$$  
The total free energy  is 
$$
    F = 8 \times \left(- \half \log \det \Delta + \log \det D_F \right) = 0 
$$
since for D=10, we have 8 transverse coordinates 
and 8 components of the unfixed  Weyl-Majorana spinor.
Thus in flat space-time the  energy associated with the supersymmetric string
is not corrected by quantum fluctuations.

\subsection{The determinant for a free BPS  quark of $AdS_5\times S^5$}

 
 The  $\kappa$ fixed GS action\cite{MatTse} is based on    treating
the target space as  the coset
 $SU(2,2|4) / (SO(1,4) \times SO(5))$.
 The  action incorporates the coupling to  the RR field. 
 The square of the  operator associated with the fermionic fluctuations is
$$
8\times \ \left( {\cal O}^2_\psi= \pa_\sigma^2 
+ \pa_\tau^2 -{3\over 4\sigma^2} \right)
$$

 The bosonic operators are  of the form
\beq
3\times \ [ {\cal O}_x&=&\pa_\sigma^2 
+ \pa_\tau^2 -{2\over \sigma^2} ]\nonumber \\
5\times \ [ {\cal O}_\theta&=&\pa_\sigma^2 
+ \pa_\tau^2 ] \nonumber \\
\eeq
where $\{x^0, x, u,\theta\}\equiv 
\{\tau, \sigma\xi_x,{1\over \sigma}, \xi_\theta\}$
and $\theta$ is the coordinate on the $S^5$
 According to a theorem of McKean and Singer\cite{MacSin} the  divergences of
a Laplacian type operator of the form 
$$\Delta= \nabla^2+X  =-{1\over \sqrt{g}} D_a(g^{ab}\sqrt{g} D_b)  +X $$
vanish if there is a match between the  fermionic and bosonic  coefficients of  
  $\nabla^2$  and $X$. In the present case  there are $8$ bosonic and $8$
fermionic $\nabla^2$ terms and hence there is no  quadratic divergences,
and the coefficients of $X$ are $8\times 3/4$ from the fermions and
$3\times 2$ from the bosons so  there are also no  logarithmic divergences. 
 It is thus clear that the divergent parts of the determinant
associated with the supersymmetric fluctuation of a BPS  string ``quark"
vanish. This problem is related to  issues associated with certain  
BPS 
soliton solutions\cite{JSV}.

\subsection{The  determinant for a Wilson line  of $AdS_5\times S^5$ }
 
 The GS action was further   simplified
by using a particular  $\kappa$ fixing\cite{KalTse}
\beq
S_{GS}&=&\int d^2\sigma  [\sqrt{g}g^{\alpha\beta} ( y^2[\pa_\alpha x^p
-2i\bar\psi\Gamma^p\pa_\alpha\psi][\pa_\beta x^p
-2i\bar\psi\Gamma^p\pa_\beta\psi]\crr
&+& {1\over y^2}\pa_\alpha y^t\pa_\beta y^t) +4\epsilon^{\alpha\beta}
\pa_\alpha y^t\bar\psi\Gamma^t\pa_\beta\psi ]\crr
\eeq
where $\psi$ is a Majorana-Weyl spinor
and the $Ads_5\times S^5$ metric is written in terms of the $4+6$
coordinates
$ ds^2 = y^2 dx_{II}^p dx_{II}^p +{1\over y^2} dy^i_{x_T}i dy^i_{x_T}$
The bosonic operators in the normal gauge now read 
\beq
2\times \qquad {\cal O}_{x_{II}}\ \ \ &=&\pa_x^2 
- {u^4\over u_0^4}\pa_t^2\nonumber \\
5\times \qquad {\cal O}_{\theta}\ \ \ &=&\pa_x^2 
- {u^4\over u_0^4}\pa_t^2 + 2{u^6\over u_0^4}\nonumber \\
1\times \qquad {\cal O}_{normal }&=&\pa_x^2 
- {u^4\over u_0^4}\pa_t^2 + 5u^2 -3{u^4\over u_0^2}\nonumber \\
\eeq
 The fermionic part of the action for the classical solution leads to the 
operator 
$$ 
    \hat{\cal O}^2_\psi = {u_0^2 \over R^2} \Gamma^1 \px + \left( {u_{cl}^4 \over {u_0^2 R^2}} 
\Gamma^0 + {u_{cl}^4 \over R^4} \cdot {\sqrt{u_{cl}^4-u_0^4} \over u_0^2} \Gamma^2 \right) \pat
$$
where we use $\Gamma$ matrices of SO(1,4), the  $AdS_5$ tangent space.
Squaring this operator, we find 
$$
    \left( {R^2 \over u_0^2} \hat{\cal O}_F \right)^2 = \px^2 - {u_{cl}^4 \over u_0^4} \pat^2  = {R^2 \over u_0^2} \hat{\cal O}_y
$$
Thus 
the transverse fluctuations ${\cal O}_{x_{II}}$ are cancelled 
by   the fermionic fluctuations. 
We are left with 6 fermionic degrees of
 freedom,  the normal bosonic fluctuation  and  5 additional bosonic
fluctuations associated with  
${\cal O}_{\theta}$.  
 Using our general result  we know that the quantum correction 
of the potential is  of a Luscher type. 
The universal coefficient and in particular  its sign  has not yet been determined.
In \cite{Stefan} the bosonic and fermionic determinants were analyzed in a different
gauge fixing procedure. It was found there that the answer is not free from 
logarithmic divergences.

\subsection{ The  determinant   for ``confining scenarios"}

Let us consider first the the setup which is dual to the pure YM thoery
in 3d. For that case
\be
f(u)  =  u^2/R^2 \qquad g(u)  =  (1 - (\frac{u_T}{u})^4)^{-1/2}
\ee
In the large $L$ limit 
 \beq
\hat{\cal O}_{x_{II}} & \longrightarrow & \frac{u_T^2}{2} \left[ \partial_x^2 +
  \partial_t^2 \right] \\
\hat{\cal O}_t & \longrightarrow & 2 u_T^2 e^{-2 u_T L} \left[ \partial_x^2 +
  \partial_t^2 \right] \\
\hat{\cal O}_n & \longrightarrow & \left[ \frac{4 u_T^2}{2R^4} +
  \frac{1}{2} \partial_x^2 + \frac{1}{2} \partial_t^2 \right]
\eeq
 We see that the operators for transverse fluctuations, 
$\hat{\cal O}_{x_{II}}$, $\hat{\cal O}_t$, 
turn out to be simply the Laplacian in flat spacetime, multiplied by overall
factors, which are irrelevant.  Therefore, the
transverse fluctuations yield the
standard L\"uscher term  proportional to $1/L$. 
 The longitudinal normal fluctuations give rise to an operator 
$\hat{\cal O}_n$ corresponding to a scalar field with mass $2 u_T/R^2 = \alpha$.
Such a field contributes a Yukawa like 
 term $$\approx -\frac{\sqrt{\alpha}e^{-\alpha L}}{\sqrt{L}}$$
to the potential. 
Thus, in the metric that corresponds to  the 
``pure YM case"
there are 7 Luscher type modes and one massive mode.  
It can be shown that a similar behavior 
occours in the general  confining setup \cite{KSSW}. 
Had the fermionic modes been those of flat space-time then the total coefficient
infront of the Luscher term would have been
a repulsive Culomb like potential\cite{GriOle} which  contradicts gauge
dynamics\cite{Bachas}. 
 However the point is that due to the RR flux 
the corresponding GS action  cannot be that of a flat space-time. Indeed
  the fermionic fluctuations  also become massive so that 
the  total inteaction is attractive after all which is in accordnce with 
\cite{DorOtt}.

\section{On the exact determination of Wilson loops}\cite{Tseytlin}
So far we have discussed the determination of Wilson loops from the classical
string description and the way  quantum fluctuations modify the classical result.
An interesting question to address is  whether in certain circomstances 
one can find an exact expression for the Wilson loop.  We start with
a string in flat space-time.  

Consider the bosonic string in flat space-time with the boundary conditions
$$X^i(\sigma=0)=0 \qquad  X^i(\sigma=\pi)=L^i  \qquad  with\ L^iL_i=L^2$$ 
 The energy 
of the lowest {\bf tachyonic} state is given by
$$
E^2=P^iP_i + m_{tach}{^2}=({L^i\over 2\pi\alpha})^2 -{(D-2)\over 24}{1\over
\alpha'} \nonumber\\
$$
so that\cite{Arvis} 
\begin{center}
\fbox
{$
E=T_{st} L\sqrt{1- {(D-2)\over 24}{1\over  T_{st}L^2}} 
 $}\end{center}
For $L>> ( T_s)^{-1/2}$ this can  be expanded to yield  
 $$\sim T_{st}L -\pi {(D-2)\over 24}{1\over L}+ ...
 $$
where the string tension $T^{-1}_{st}=2\pi\alpha'$.
Thus this expansion yields the Luscher  quadratic fluctuation term. 
 Moreover, this result is identical to the expression  for the
NG action   derived 
for a bosonic string in flat space-time  in the large
$D$ limit\cite{Alvarez}
$$ D\rightarrow \infty \qquad {\pi \over 24 T_{st}L^2}\rightarrow 0\qquad
{D\pi \over 24 T_{st}L^2}\rightarrow finite$$
It is straightforward to realize that    for a static classical configurations
$ E_{Poly}= S_{NG}$.

A more challenging question is whether one can 
 find such ``exact" solution for a non-flat spece-time.
A naive conjecture is that for the $Ads_5\times S^5$
the result is $\sim-{\sqrt{g^2N}\over L}\sqrt{1+{c\over \sqrt{g^2N}}}$.
However, whereas its large $g^2N$ expansion  
 includes the  lresult of \cite{Mal2} 
and a non trivial Luscher term, it does not permit a smooth extrapolation
to the weak coupling region where  the potential behaves like 
$\sim -{g^2N \over L}$. 
\subsection{ Wilson loops for string actions with WZ term}
In general  exact results are known  for non trivial backgrounds  of
 group manifolds and coset spaces.
The  sigma model associated with such target spaces is equipped with  a WZ
term 
The bosonic action is therefore 
$$S_B = S_{NG} + \int d^2 \sigma e^{\alpha\beta} \pa_\alpha X^\mu 
\pa_\beta X^\nu B_{\mu\nu}$$

For the case that  the only non-trivial component of $ B_{\mu\nu}$
is  $B_{01}=B(u)$  one finds that for $B\neq f$ ( $f$ was defined in (2))
$$
S_{NG+WZ} =\int_{u_0}^\infty du
{g\over f}{f^2-B(f_0+B-B_0)\over \sqrt{f^2-(f_0+B-B_0)^2}}
$$

and  
$$ S_{NG+WZ} = 0 \qquad for\   B=f$$

For the former case 

$$
S=(f_0+B_0)L +2\int_{u_0}^\infty du 
{g\over f} \sqrt{f^2-(f_0+B-B_0)^2}
$$
For sting  theories where $f_0+B_0$, which is the value of $B+f$ at the 
minimum of the string configuration, 
 does not vanish  the WIlson loop
admits an area law behavior  with  a string tension  equal to $f_0+B_0$.

In the string model associated with a three dimensional 
 $SL(2,R)$ group manifold, and in the ``near extremal" corresponding models of 
 ${SL(2,R)\over R}\times R$\cite{HorTse}. 
The $B$ term match the $f$, namely $f=B$ so that 
the Wilson line is a straight line and th energy $E=0$.




{\bf Acknowledgements}
I would like to thank Y. Kinar, E. Schreiber, A Tseytlin and N. Weiss 
for useful discussions. 
Research supported in part by 
the US-Israeli Binational Science Foundation,GIF, 
the German-Israeli Foundation for Scientific Research  and the Israel Science
Foundation.


\vskip 1 cm  





\begin{thebibliography}{99}

\bibitem{Mal2} J. Maldacena, "Wilson loops in large $N$ field theories", Phys. Rev. Lett. 80 (1998) 4859-4862, hep-th/9803002.

\bibitem{Rey} S.-J. Rey and J. Yee,  ``Macroscopic strings as
    heavy quarks in the large N gauge theory and anti-de Sitter
    supergravity'', hep-th/9803001.
\bibitem{BISY1} A. Brandhuber, N. Itzhaki, J. Sonnenschein, S. Yankielowicz, 
"Wilson Loops in the Large N Limit at Finite Temperature", hep-th/9803137.

\bibitem{RTY} S.J. Rey, S. Theisen and J.T. Yee, "Wilson-Polyakov loop
at finite Temperature in large $N$ gauge theory and anti-de Sitter 
supergravity", hep-th/9803135.

\bibitem{BISY2} A. Brandhuber, N. Itzhaki, J. Sonnenschein, 
S. Yankielowicz, "Wilson Loops, Confinement,
 and Phase Transitions in Large N Gauge Theories from Supergravity", 
JHEP 9806 (1998) 001,hep-th/9803263

\bibitem{GroOog}  D. Gross and H. Ooguri
``Aspects of Large N Gauge Theory Dynamics as Seen by String Theory"
 Phys.Rev. D58 (1998) 106002,  hep-th/9805129
\bibitem{GriOle}  J. Greensite, P. Olesen,
 "Remarks on the Heavy Quark Potential in the Supergravity Approach", 
hep-th/9806235;
"Worldsheet Fluctuations and the Heavy Quark Potential 
in the AdS/CFT Approach", hep-th/9901057.
\bibitem{DorOtt} H. Dorn and   H.-J. Otto
``On Wilson loops and $Q\bar Q$-potentials from the AdS/CFT 
relation at $T\geq 0$"  hep-th/9812109;
H. Dorn, V. D. Pershin `` Concavity of the $Q\bar Q$ potential in ${\cal N}=4$ super Yang-Mills gauge theory and
              AdS/CFT duality" hep-th/9906073

\bibitem{gppz}  L. Girardello, M. Petrini, M. Porrati and A. Zaffaroni,
   ``Confinement and Condensates Without Fine Tuning in
    Supergravity Duals of Gauge Theories'', hep-th/9903026.


\bibitem{KSS}  Y. Kinar, E. Schreiber and J. Sonnenschein,
{\em ``$Q \bar{Q}$ Potential from Strings in Curved Spacetime -
  Classical Results''}, hep-th/9811192.
\bibitem{DGO} N. Drukker, D. Gross and H. Ooguri
 "Wilson Loops and Minimal Surfaces", hep-th/9904191 
\bibitem{IMSY}  N. Itzhaki,  J. Maldacena, 
J. Sonnenschein and  S. Yankielowicz, "Supergravity And The Large $N$
Limite Of Theories With Sixteen Supercharges", hep-th/9802042.



\bibitem{Witads2} E. Witten, "Anti-de Sitter Space, Thermal Phase
    Transition, and Confinement in Gauge Theories", hep-th/9803131.
             





\bibitem{KSS1} Y. Kinar, E. Schreiber, J. Sonnenschein, "Precision 'Measurements' of the $Q\bar{Q}$ Potential in MQCD", hep-th/9809133.
\bibitem{Luscher}  M. Luscher, K. Symanzik, P. Weisz, 
"Anomalies of the free loop wave equation in the WKB approximation"
Nucl. Phys. B173 (1980) 365.


\bibitem{CORT} C. Cs\'aki, Y. Oz, J.Russo, J. Terning, "Large $N$ QCD from Rotating Branes", hep-th/9810186.


\bibitem{LuiTse}  Hong Liu, A. A. Tseytlin
``D3-brane - D-instanton configuration and N=4 super YM theory in constant self-dual
              background"
Nucl.Phys. B553 (1999) 231-249,hep-th/9903091
\bibitem{Li} M. Li, {\em ``'t Hooft Vortices on D-branes''},
   JHEP 9807 (1998) 003, hep-th/9803252.

\bibitem{KSSW} Y. Kinar, E. Schreiber, J. Sonnenschein, N.Weiss
 to appear
\bibitem{MatTse} R. R.  Matsev and A. A. Tseytlin
``Supersymmetric D3 brane action in $AdS_5 x S^5$"
Phys.Lett. B436 (1998) 281-288, hep-th/9806095
\bibitem{KalTse} R. Kalosh and A. A. Tseytlin,
``Simplifying superstring action on $AdS_5 x S^5"$  
JHEP 9810 (1998) 016, hep-th/9808088
\bibitem{MacSin} McKean and I. Singer  J. Diff. Geom. 1 (1967) 43.
\bibitem{JSV} N. Graham, R. L. Jaffe 
Phys.Lett. B435 (1998) 145-151,  hep-th/9805150, 
Nucl.Phys. B544 (1999) 432-447, hep-th/9808140;

M. Shifman, A. Vainshtein, M. Voloshin,
Phys.Rev. D59 (1999) 045016 hep-th/9810068;

H., M. Stephanov, P. van Nieuwenhuizen, A. Rebhan
Nucl.Phys. B542 (1999) 471-514 hep-th/9802074

\bibitem{Stefan} S. Forste, D. Ghoshal and S. Theisen 
Stringy Corrections to the Wilson Loop in N=4 Super Yang-Mills Theory",
 JHEP 9908 (1999) 013,       hep-th/9903042
\bibitem{Bachas} C. Bachas,  Phys. Rv. D33 (1986) 2723.
\bibitem{Tseytlin} This section is based on discussions with A.A. Tseytlin
\bibitem{Arvis} J. F. Arvis Phys. Lett. 127B (1983) 106.
\bibitem{Alvarez}O. Alvarez, Phys. Rev. D24 (1981) 440. 
\bibitem{HorTse}G. Horowitz and A. A. Tseytlin 
``A new class of exact solutions in string theory"
hep-th 9409021




















\end{thebibliography}
\end{document}